\documentclass[twocolumn,english]{revtex4-1}
\usepackage[T1]{fontenc}
\usepackage[latin9]{inputenc}
\usepackage{xcolor}
\usepackage{amsbsy}
\usepackage{amstext}
\usepackage{amssymb}
\usepackage{graphicx}
\usepackage{wasysym}
\usepackage{esint}
\PassOptionsToPackage{normalem}{ulem}
\usepackage{ulem}

\makeatletter

\providecolor{lyxadded}{rgb}{0,0,1}
\providecolor{lyxdeleted}{rgb}{1,0,0}

\DeclareRobustCommand{\lyxsout}[1]{\ifx\\#1\else\sout{#1}\fi}

\makeatother

\usepackage{babel}
\begin{document}

\title{Testing the physics of knots with a Feringa nano engine}

\author{M. Lang,$^{1}$ C. Schuster,$^{1,2}$ R. Dockhorn,$^{1,2}$ M. Wengenmayr,$^{1,2}$
J.-U. Sommer$^{1,2}$ }

\address{$^{1}$Institut Theorie der Polymere, Leibniz-Institut für Polymerforschung
Dresden e.V., Hohe Straße 6, 01069 Dresden, Germany}

\affiliation{$^{2}$Technische Universität Dresden, Institute for Theoretical
Physics, Zellescher Weg 17, 01069 Dresden, Germany }
\email{lang@ipfdd.de}

\begin{abstract}
We use the bond fluctuation model to study the contraction process
of two polymer loops with $N$ segments that are connected each to
the bottom and top part of a Feringa engine. The change in the size
of the molecules as well as the folding of the two strands follows
approximately scaling predictions that are derived by assuming that
the strands are confined inside an effective tube. Conformation data
can be overlapped when plotting it as a function of $W_{\text{n}}N^{-1/4}$,
where $W_{\text{n}}$ is the winding number of the two strands that
is proportional to the number of blobs inside the ``knotted'' region
of the molecule and $N$ is the degree of polymerization of the strands.
Our data supports a weak localization of the knots along the contour
of flexible cyclic a-thermal polymers with a localization exponent
$t\approx0.78$.
\end{abstract}
\maketitle

\section{Introduction}

The physics of knotting is relevant for any kind of long flexible
molecule that is free to select its conformations. Knots affect the
size of the molecules \cite{Moore2005}, equilibrium and non-equilibrium
dynamics \cite{Mai2016,Klotz2018}, deformation behavior \cite{Caraglio2015},
translocation through a pore \cite{Suma2015,Suma2017}, or protein
folding \cite{Ziegler2016} to provide just some examples. One key
question is whether knots self-tighten for entropic reasons or not
\cite{Grosberg2016}, as this self-tightening controls the portion
of the molecule that is subject to a knotted conformation. 

This question originates from physics motivated models for the entropy
of a knotted molecule, for instance ref. \cite{Grosberg1996}. Here,
the self-confinement of the chains is modeled by considering a maximally
inflated confining tube with the same knot topology as the chain inside
the tube. Inflating the tube to a diameter $\xi$ until it arrests
at a contour length $L$ results in a weak topological invariant $p=L/\xi$
of tube segments equivalent to the ``ropelength'' \cite{Cantarella1996}
used by mathematicians to characterize different ``ideal'' knots
\cite{Katrich1996}. For a chain consisting of $N$ statistical segments
of length $b$, one way to realize optimum conformations is \cite{Grosberg2000}
within such a maximum inflated tube where the maximum entropy $S$
of the embedded chain is found by balancing Pincus tension blobs with
compression blobs. As a result, the size of the molecule, $R$, varies
as $R\propto p^{-1/6}$ \cite{Grosberg2000}. Another way is to assume
a ``phase separation'' between a tightly knotted section of $\approx p$
segments and an unconstrained loop of roughly $\approx N-p$ segments
\cite{Grosberg2000}. In the latter case, the resulting molecular
size $R\propto p^{-\nu+1/3}$ is also a function of the rope length
$p$, whereby $\nu\approx0.5876$ \cite{Clisby2010} is the Flory
exponent. ``Phase separation'' (self-tightening of the knot) is
expected, if $p\ll N^{0.2}$, while the knot spreads over the whole
polymer otherwise. As the coefficients for the phase boundary are
not known, one cannot rule out that extremely long chains with $N$
on the order of $10^{6}$ or even much larger might be necessary to
observe phase separated knots, since the smallest possible ropelength
of the trefoil knot is $p\approx16.4$ \cite{Katrich1996} and there
is no way to reduce $p$ below this minimum for a fixed knot topology.

Several simulation studies have addressed this question in the past
\cite{Marcone2005,Marcone2007,Farago,Dai2015} but arrive at different
results concerning the power $t$ that describes the localization
of the knotted section within $N^{t}$ segments. For closed random
walks, $t=0$ was obtained in refs. \cite{Katritch2000,Millet2011}.
For self-avoiding walks, an exponent of $t=0.4\pm0.1$ \cite{Farago}
or $t\approx0.75$ \cite{Marcone2005,Marcone2007} was obtained using
different approaches. Recent data by Dai et al. \cite{Dai2015} on
the knotting of linear self-avoding chains were considered to be more
in line with a preferred knot size independent of $N$, i.e. $t=0$,
but it was suspected that still a fat tail could affect the scaling
of the average size of a knot \cite{Grosberg2016}.

A general problem for analyzing knot localization is that real polymers
exhibit force extension relations \cite{Rubinstein2005} quite different
to a Gaussian coil or a self-avoiding chain when the tube diameter
is getting squeezed to a size comparable to individual segments \cite{Kim2013}.
Thus, significant corrections to the free energy estimates in ref.
\cite{Grosberg1996} might be necessary. On top of that, the algorithm
for analyzing the knotted section could lead to artifacts \cite{Tubiana2011b},
in particular for more complex knots or collapsed conformations of
a knot. Beyond that, additional finite $N$ corrections come into
play when the molecule consists not of much more tension blobs as
the ropelength of the knot, see the Appendix C for more details. Taking
these points together, self-tightening of knots still appears to be
a Gordian knot.

Recent work where rotating molecular engines were attached to polymers
\cite{Li2015,Weysser2015} possibly provides the sword to untie this
Gordian knot. These nano engines allow to tune the rope length $p$
of a figure of 8-shaped ``tanglotron'' molecule (``T8'', see Figure
\ref{fig:Tanglotron-molecule-with}.) continuously down to zero, which
reduces dramatically the required $N$ to observe a possible phase
transition between self-tightened and spreaded state. The resulting
topology of the T8 under the action of the engine ist not truely a
knot and it does not refer exactly to a linked state of two cyclic
polymers, since the central unit of the engine is located on both
cyclic strands. But this is unproblematic for our analysis, since
the similarity of linked and knotted structures concerning knot localization
has been demonstrated in previous works \cite{Metzler2002,Baiesi2011}
and the common point induces corrections only of order $1/N$. Actually,
the original work by Grosberg \cite{Grosberg1996} ignores explicitly
such details for the sake of a general physical model that applies
independent of the type of knot or link.

In our work, we discuss first our simulations before we analyze the
properties of the T8 in the limit of small $p$ concerning signatures
of self-tightening or a phase transition from a spreaded to non-spreaded
state. We further use the linear deformation regime at small applied
torques $M$ of the molecular engine to measure the topological potential
that unfavours the formation of a larger rope length $p$. This provides
an alternative access to the localization exponent, whereby the results
can be double checked by the conformation changes as a function of
$p$ without need for an algorithm to detect knot localization. This
last point is of particular relevance for our study, since the known
algorithms to identify the knotted part may lead to increasingly ambiguous
results for more complex knots or more collapsed conformations of
a knot \cite{Tubiana2011b}. In our simulations, the complexity of
the link increases with torque and enforces a collapse of the molecule.
Thus, a systematic bias when using one of these methods to detect
knot localization could not be excluded, but this difficultes are
avoided with our approach.

\section{Simulation Details and Analysis}

For simulation, we use the bond fluctuation model (BFM) by Carmesin
and Kremer \cite{Carmesin1988} and Deutsch and Binder \cite{Deutsch1991}.
It is a Monte Carlo method to simulate universal properties of polymers
on a lattice. In this model, a monomer occupies a cube of 8 adjacent
lattice sites on a simple cubic lattice. The monomers are connected
by bond vectors, which are restricted to a specific set of 108 vectors
of length between $2$ and $\sqrt{10}$. These vectors are defined
such that a test of excluded volume for a motion to one of the nearest
6 lattice positions is sufficient to preserve topology, if the bond
vectors from the new position are still contained in the set of allowed
bond vectors. Monomers and motion directions are chosen randomly.
A Monte-Carlo step consists of $m$ attempted monomer moves, where
$m$ is the number of monomers in the simulation box. 

\begin{figure}
\begin{center}\includegraphics[width=0.94\columnwidth]{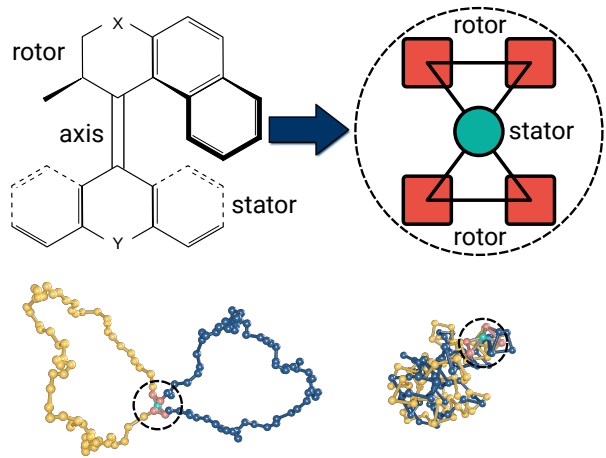}\end{center}

\caption{\label{fig:Tanglotron-molecule-with}Left top: chemical structure
of a second generation Feringa engine \cite{Koumura2002}. Right top:
implementation as a double rotor molecule within the simulations.
Bottom: unfolded state of the T8 with a nano engine at its core (left)
and completely coiled state (right). See also Ref. \cite{supporting}
for a short movie that shows the coiling of the structure in response
to the action of the engine. All Figures color online.}
\end{figure}

The Feringa engine is built up of two parts called rotor and stator,
see top left part of Figure \ref{fig:Tanglotron-molecule-with}. The
rotor undergoes conformation changes in the way that it rotates unidirectional
around the stator when illuminated with light \cite{Koumura1999,Koumura2002}.
To model a similar qualitative behavior with the BFM, the motor is
made of five monomers as shown in the top right part of Figure \ref{fig:Tanglotron-molecule-with}.
At X and Y, the Feringa engine is connected to the surrounding structure;
in the simulations this happens on all rotor monomers in order to
cause a twist of the connected pair of chains. For illustration, Ref.
\cite{supporting} contains a short movie that shows how the Feringa
engine coils up the two attached polymer strands.

In a lattice based Monte-Carlo simulation, it is not feasible to completely
block the backwards rotation of the rotors as for the original Feringa
engine, since this could cause a freezing of the engine. Thus, we
use the Metropolis algorithm \cite{Metropolis1953} to introduce a
penalty for backwards rotation based upon the potential energy difference
\begin{equation}
\Delta U=\pm M\Delta\alpha\label{eq:DeltaU}
\end{equation}
between new and old position of the particle. Here, $M$ is an effective
torque acting on the rotor to model the rotational engines of refs
\cite{Koumura1999,Koumura2002}. $\Delta\alpha$ is the change in
the angle of twist between upper and lower rotor upon moving a monomer
and the sign in front of $M$ determines whether the engine drives
right handed (negative) or left handed (positive) twist. Note that
a proper computation of $\Delta\alpha$ leads to some non-trivial
restrictions for our lattice based simulation model, which are discussed
in the Appendix A. The motion of a rotor monomer is performed only
with probability $w$ 
\begin{equation}
w=\min\left[1,\exp\left(-\frac{\Delta U}{kT}\right)\right].\label{eq:Metropolis}
\end{equation}

Note that this implementation is qualitatively similar to the function
of a Feringa engine during irradiation for an appropriate choice of
$M$ and at low total twist of the attached polymers, but not after
the light has been switched off or in the limit of high torque. The
Feringa engine still preserves the torsion angle between upper and
lower rotor, while setting $M=0$ in the simulations either results
in a backwards rotation (if potential energy was stored in the attached
polymer structure) or a random drift of the torsion angle. This could
be adapted in the simulations by inserting an additional bond that
connects both rotors of the engine and blocks further rotation at
the end of the irradiation. Furthermore, the increase of the torsion
angle is not monotonic in the simulations and fluctuations $>2\pi$
around an average torsion angle are always possible, which is not
the case for the original Feringa engine. 

Note that the gels with Feringa engines inside were breaking after
a long time exposure with light in ref. \cite{Li2015}. To allow for
a better control of the induced shrinking process and a reversible
use of the materials, either a second mechanism to unwind the chains
is required or the development of engines that control torque as a
function of light intensity. The former possibility has been realized
recently \cite{Foy2017}. Related research by one of the authors of
ref. \cite{Li2015} already considers equilibrium physics simulations
to model the action of such engines \cite{Weysser2015}. Our research
is in line with this modification of the engines to make them more
suitable for polymer applications.

The choice of the BFM as simulation model in contrast to molecular
dynamics as in ref. \cite{Weysser2015} is motivated by the high performance
of the BFM that allows to scale up to gels containing Feringa engines.
Certainly, we cannot address the dynamic response of the attached
molecules to the action of the engine with our lattice based simulations.
But Monte-Carlo methods are known to be very effective tools to study
universal properties related to molecular conformations at equilibrium
\cite{Binder2008}, which is the task of our investigations.

The typical response of a single attached flexible polymer strand
to the action of a Feringa engine is not of particular interest, since
the action of the engine is typically much slower than the thermal
(torsional) motion of the attached monomers. This is different to
double stranded DNA with an intrinsic torsional stiffness, where an
applied torque can make a single ds-DNA strand to coil up with itself
\cite{Strick1996}. For flexible polymers, a significant response
is achieved when at least two strands are attached to either side
of the Feringa engine, which now causes these strands to wind around
each other. This type of coiling is most conveniently measured by
summation over all changes of the angle of twist $\Delta\alpha$ of
a given engine, which provides the change in the winding number \cite{Ricca2009}
\begin{equation}
\Delta W_{\text{n}}=\frac{\sum\Delta\alpha}{2\pi}
\end{equation}
between pairs of strands connecting from upper to lower rotor upon
the action of the engine. 

Alternatively, one can consider the linking number 
\begin{equation}
L_{\text{k}}=\frac{1}{4\pi}\oint_{x_{1}}\oint_{x_{2}}\vec{x}_{1}'(s_{1})\times\vec{x_{2}}'(s_{2})\cdot\frac{\vec{x}_{1}(s_{1})-\vec{x_{2}}(s_{2})}{\vert\vec{x}_{1}(s_{1})-\vec{x_{2}}(s_{2})\vert^{3}}\text{ d}s_{1}\text{ d}s_{2},\label{eq:Lk}
\end{equation}
as defined for two closed non-crossing curves $x_{1}$ and $x_{2}$
and their tangent vectors $\vec{x}'(s)$ at $s$. Here, still one
needs to connect monomers A and C as well as B and D such that curve
closure does not lead to an intersection of the two curves \cite{Adams1995}.
Note that in contrast to the winding number, the Linking number is
always an integer. Since $W_{\text{n}}-1\le L_{\text{k}}\le W_{\text{n}}+1$
when starting from uncoiled conformations of the 8-shaped tanglotron
(``T8''), we used this dependence to double check our computation
of $W_{\text{n}}$ and $L_{\text{k}}$. For convenience, however,
the angle of twist was measured continuously to determine the average
winding numbers $W_{n}$ at a given torque. 

\section{``Knot'' localization}

Let us consider first simple scaling arguments to understand the behavior
of the \emph{maximum inflated tube} (spreaded knot). In our simulations,
we apply a stochastic force $f_{\text{t}}$ at a distance $b/2$ from
the rotation axis of the engine such that the applied torque at the
engine is $M\approx f_{\text{t}}b/2.$ The action of the engine coils
up the two strands and we expect a tight double helix conformation
of tension blobs for maximum inflated tubes similar to ref. \cite{Weysser2015}.
For a tube diameter (blob size) $\xi>b$ (or equivalently an applied
torque of roughly $M\lesssim kT$), the force $f_{\text{t}}$ is leveraged
by excluded volume interactions within the first blob to a distance
of $\approx\xi$ instead of $\approx b$ from the axis of the helix
of tubes. The pulling force along the tube is not modified, since
the repulsion of the strands that drives the leverage is orthogonal
to $f_{\text{t}}$ (as a mechanical analogue one could consider a
pulley at distance $\xi/2$ and pulling the chain with force $f_{\text{t}}$
through this pulley). This leads for $M\lesssim kT$ to an effective
torque $M_{\text{e}}\approx M\xi/b\approx kT$ inside the entangled
section that is larger than the ``applied'' torque $M$ by a factor
of $\xi/b$. Thus, instead of controlling torque, the action of the
tanglotron sets the tube diameter $\xi\approx bkT/M\approx kT/f_{\text{t}}$.
Using the scaling relation between the number of segments per blob
and blob size in good solvents, $g\approx\left(\xi/b\right)^{1/\nu},$
one can estimate, therefore, the change in free energy by counting
the number of tension blobs
\begin{equation}
\Delta F\approx kT\frac{N}{g}\approx kTN\left(\frac{M}{kT}\right)^{1/\nu}\label{eq:F3}
\end{equation}
as a function of the applied torque $M$.

It has been shown that optimum packing of a double helix (minimal
tube length $L$ at a given tube diameter and winding number $W_{\text{n}}$)
is obtained for $L\approx W_{\text{n}}\xi\left(\pi^{2}+4\right)^{1/2}$
\cite{Weysser2015}. Since the cost in free energy for chain stretching
is a known function of $L$ \cite{Pincus1976}, we arrive at a linear
relation between change in free energy and the winding number:
\begin{equation}
\Delta F\approx\left(L/\xi\right)kT\approx pkT\approx W_{\text{n}}\left(\pi^{2}+4\right)^{1/2}kT.\label{eq:DF-1}
\end{equation}
In combination with the previous estimate of free energy, this yields
\begin{equation}
W_{\text{n}}\propto NM^{1/\nu}.\label{eq:Wn-1}
\end{equation}

\begin{figure}
\includegraphics[angle=270,width=1\columnwidth]{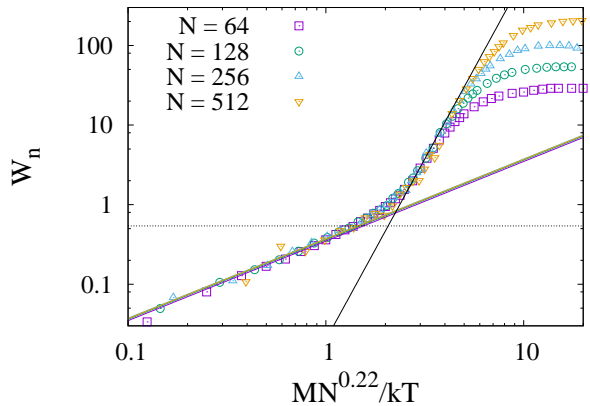}

\caption{\label{fig:Winding-number-as}Winding number as a function of the
scaled torque. The dotted line is at a level of $W_{\text{n}}=2/\left(\pi^{2}+4\right)^{1/2}$
where there is in average one blob per strand inside the helix. The
colored continuous lines are linear fits to the data at $MN^{0.22}<1$.
Overlap of the data was optimized with respect to the smallest variance
of the coefficients of these linear fits. The black continuous line
is the relation $W_{\text{n}}\propto M^{1/(1-t)}$ with $t=0.78$.}
\end{figure}

Let us now develop a simple model for the \emph{phase separated state}.
Here, we assume that an entropic force is pulling a significant fraction
of chain segments out of the entangled zone. Following the above discussion,
we argue that the entropic force defines the number of segments per
tension blob, $g$, inside the entangled zone as there is $\xi>b$.
In order to check for a possible weak localization, we write $g\propto N^{t}$
with a localization exponent $t$ to be determined from the simulation
data. In the limit of low applied torque $M\ll kT$, we expect a linear
force extension relation $M\propto W_{\text{n}}$ up to the point
where pulling force $f_{\text{t}}$ and entropic force compensate
each other similar to the initial linear regime of a Pincus chain
\cite{Pincus1976}. Coiling of the two strands occurs for torques
larger than this threshold. The free energy cost for stretching an
equivalent linear chain to $g$ segments per tension blob is $\approx N/gkT\approx N^{1-t}kT$.
Since there must be $M\propto N/g$ for the linear regime, we determine
the exponent $t$ by re-scaling $M$ with a factor of $N^{1-t}$ such
that the $W_{\text{n}}$ data for small $M$ collapse on top of each
other. This re-scaling of the linear regime works well as demonstrated
in Figure \ref{fig:Winding-number-as} and leads also to a collapse
to the data in the following non-linear regime up to the saturation
point where the data levels off depending on $N$. Optimum collapse
is obtained for $t\approx0.78$, which is almost the same as in refs.
\cite{Marcone2005,Marcone2007} with $t\approx0.75$, but does not
agree \footnote{The force deformation data in Fig. 2 of ref. \cite{Farago} do not
scale as expected; it is therefore not clear whether the simulation
data in ref. \cite{Farago} are consistent or not.} with the $t=0.4\pm0.1$ obtained in ref \cite{Farago} or the $t\approx0$
of \cite{Dai2015}.

The spreaded knot regime is not truly visible in Figure \ref{fig:Winding-number-as}
and could be hidden in the transition to saturation. Another possibility
is that the T8 switches between a spreaded and a phase separated state
within the regime where $W_{\text{n}}$ grows rapidly, since the free
energies of these state could be of the same order of magnitude for
the limited $N$ of our study (the free energy penalty grows apparently
just as $N^{0.22}$). In consequence, we would expect a significant
jump in the winding numbers when switching from a phase separated
to spreaded knot state, since the nano engine just triggers the blob
size. 

\begin{figure}
\includegraphics[angle=270,width=1\columnwidth]{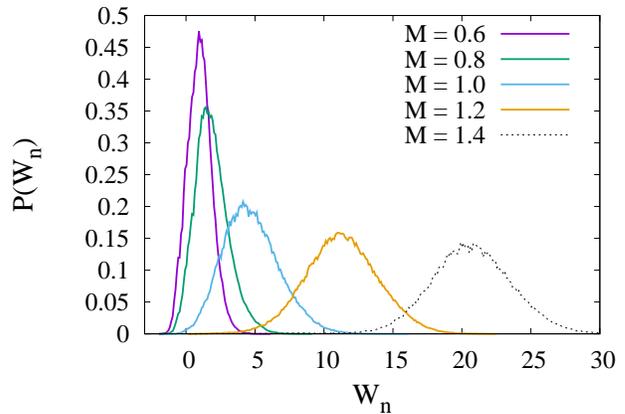}

\caption{\label{fig:Distribution-of-winding}Normalized distribution of winding
numbers for a given applied torque $M$ and $N=250$. The average
winding numbers for $M=0.6,$ $0.8,$ $1.0,$ $1.2,$ and $1.4$ are
$W_{\text{n}}\approx0.99$, $1.99$, $4.68$, $10.7$, and $20.4$. }
\end{figure}

In order to scan for such a smeared out discontinuous phase transition,
we analyzed the distribution of winding numbers, $P(W_{\text{\ensuremath{\ensuremath{n}}}})$,
for all torques and examples are shown in Figure \ref{fig:Distribution-of-winding}.
Our data shows no indication for two competing ground states with
two separate peaks in $P(W_{\text{\ensuremath{\text{n}}}})$ such
that for the range of parameters of our study, the transition is a
continuous one. As a complementary test, we also analyzed the contacts
between segments on the two loops (see Appendix B). The data of this
analysis is in accord with a localization of the entanglements next
to the Feringa engine for torques $M\lesssim1.2$ (prior to the onset
of saturation). Therefore, we conclude that there is a weak localization
of the entanglements next to the engine with a localization exponent
$t\approx0.78$ as discussed above. This weak localization covers
essentially all of the accessible parameter space between initial
linear and final saturation regime.

One striking feature of the data in Figure \ref{fig:Distribution-of-winding}
is the broad distribution of $W_{\text{n}}$ for a given torque and
correspondingly, the number of blobs in the entangled zone. These
fluctuations smear out the folding transition. In consequence, possible
corrections to scaling can hardly be extracted from the data. Thus,
we attempt only a very simplistic description of the regime where
$W_{\text{n}}$ increases strongly. Let us assume that this growth
occurs in self-similar manner. The corresponding power has to settle
an increase of $W_{\text{n}}$ $\propto N$ for a window of torques
$M\propto N^{1-t}$, which results in the proposal that a very rough
approximation might be obtained by a power law
\begin{equation}
W_{\text{n}}\propto M^{1/(1-t)}\approx M^{4.55}.\label{eq:Wn}
\end{equation}
This simplistic estimate is included into Figure \ref{fig:Winding-number-as}
and agrees surprisingly well with the simulation data.

\section{Folding transition}

\begin{figure}
\includegraphics[angle=270,width=1\columnwidth]{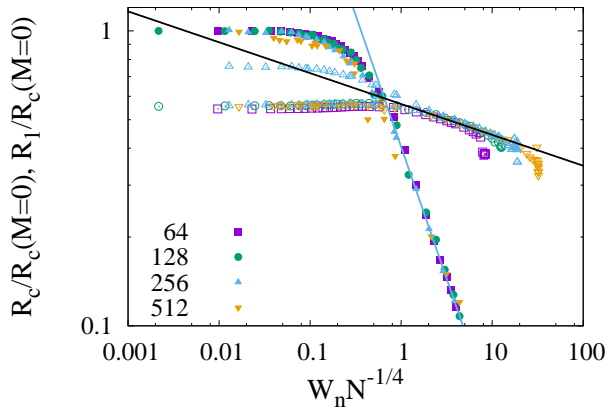}

\caption{\label{fig:Folding-of-the}The center to center distance of both loops,
$R_{\text{c}}$, (full symbols) the gyration radius of a single loop,
$R_{1}$, (lower set of data with open symbols) are shown as a function
of the scaled winding number $W_{\text{n}}N^{-1/4}$. For comparison,
the gyration radius of the whole molecule, $R_{\text{g}}$, is shown
for one particular sample ($N=256$, open triangles). All data are
normalized with respect to $R_{\text{c}}$ at zero torque, $M=0$.
The lines are power law fits for large $W_{\text{n}}$ yielding $-0.89\pm0.03$
for the $R_{\text{c}}$ data and $-0.10\pm0.01$ for the $R_{1}$
data respectively.}
\end{figure}

The folding process of the T8 is monitored best by the distance between
the centers of mass of both loops, $R_{\text{c}}$, since it shows
the strongest change as a function of $W_{\text{n}}$, see Figure
\ref{fig:Folding-of-the}. Recall that $p\propto W_{\text{n}}$ and
that the position of the transition point should scale as $p\approx N^{0.2}$.
Thus, we expect a collapse of all data, when plotting it as a function
of $W_{\text{n}}N^{-0.2}$. As above, we vary the power of $N$ to
scan for optimum collapse of the data, which is found for $W_{\text{n}}N^{-1/4}$.
This is close to the original prediction, which strongly supports
the model of Grosberg et. al. \cite{Grosberg1996,Grosberg2000}.

Let us discuss now the scaling of the conformation changes of the
T8. Figure \ref{fig:Folding-of-the} shows that distinct power laws
can only be tested for $W_{\text{n}}N^{-1/4}>1$, which refers to
the spreaded knot regime. As ``folding point'' we identify the intersection
of the high $W_{\text{n}}$ scaling regimes that almost coincides
with the condition that the center-to-center distance equals the gyration
radius of a single loop, $R_{\text{c}}=R_{1}$. One key parameter
for our scaling analysis is the knotting length of a cyclic polymer,
$N_{0}$, which is in the range of 3000 segments for the bond fluctuation
model in melt \cite{Michalke2001}, while it can be expected in the
range of $10^{5}$ segments for isolated molecules in good solvent
\cite{Koniaris1991}. Therefore, our data is clearly in the limit
of $N<pN_{0}$, for which a shrinkage of the size of the molecule
with increasing complexity of the knot, $R\propto p^{-1/6}$, was
proposed previously \cite{Grosberg2000}. We obtain $R\approx R_{1}\propto p^{-0.10\pm0.01}$,
which is a smaller power than predicted. However, the range of data
$W_{\text{n}}N^{1/4}>1$ is still quite narrow to truly test a weak
power law $p^{-1/6}$. Furthermore, for increasing $W_{\text{n}}$,
our samples gradually cross over to the regime where the chain sections
inside the blobs are overstretched. Overstretched chains show a sub-linear
response to an applied force, which results in a smaller apparent
power for the change in $R$. This explains qualitatively the observed
discrepancy to the theoretical prediction.

In order to discuss the scaling of $R_{\text{c}}$ in the folded state,
we assume a helix like conformation of the effective tubes where the
chains are confined similar to ref. \cite{Weysser2015} and in agreement
with the model of Grosberg \cite{Grosberg1996,Grosberg2000}. Under
these conditions, chain conformations are modeled by $\propto W_{\text{n}}$
blobs of size $\xi\propto bg^{\nu}\propto b\left(N/W_{\text{n}}\right)^{\nu}$.
Each of the $\propto W_{\text{n}}$ pairs of blobs along the helix
provides an independent measurement for the distance $\xi$. We interpret
the scaling of the center of mass distance $R_{\text{c}}$ as a series
of $\propto W_{n}$ independent measurements of the inter-blob distance,
thus, we expect $R_{\text{c}}\approx\xi W_{\text{n}}^{-1/2}\propto bN^{\nu}W_{n}^{-(\nu+1/2)}\propto W_{\text{n}}^{-1.08}$.
Our simulation data fits best to $W_{\text{n}}^{-0.89\pm0.03}$, which
is again a somewhat weaker dependence as expected that could be explained
qualitatively by the same reasons as above. Altogether, also the conformation
changes support the model of Grosberg et al. \cite{Grosberg1996,Grosberg2000}.

\section{Discussion}

The above results allow for a distinct view on previous simulation
studies, in particular concerning the different results for the localization
exponent ranging from $t\approx0$ \cite{Dai2015} over $t\approx0.4$
\cite{Farago} to $t\approx0.75$ \cite{Marcone2005,Marcone2007}.
The major differences among these works - beyond ``boundary effects''
for cyclic vs. linear polymers - are the existence of an external
pulling force (only in ref. \cite{Farago}) and the algorithm to detect
knot localization.

With our data at hand, we cannot judge on the impact \cite{Tubiana2011b}
of the different algorithms that were used. On the contrary, we have
to stress that such a discussion does not affect our results, as we
determine the localization potential directly by analyzing the entropic
cost to create additional ``blobs'' for increasingly coiled conformations
of two intertwined polymer strands. The analysis in ref. \cite{Dai2015}
yields an effective tube diameter of roughly three times the size
of a monomer, which is within the regime where classical scaling laws
no longer lead to a reasonable description of cylindrically confined
chains \cite{Kim2013}. Together with the suspection that still a
fat tail could affect the scaling of the average size of a knot \cite{Grosberg2016}
and our discussion in the Appendix C, it remains difficult to conclude
towards a yet asymptotic behaviour. With respect to the above discussion,
it is indeed rather surprising that our data agrees well with refs.
\cite{Marcone2005,Marcone2007} on cyclic polymers also concerning
the still limited ratio of $N/(gp)$ for a fixed knot topology.

Recent work on the active supercoiling of DNA \cite{Racko2015} investigates
also the active coiling of molecules by means of computer simualtions,
here the dynamic coiling of individual DNA strands and the relevance
of this problem for the unknotting and postreplicative decatenation
of DNA. Research in this direction, however, would require to use
non-equilibrium molecular dynamics simulations. With our tools at
hand, the investigation of more complex systems at equilibrium is
straigth forward to address. Therefore, our current interest focusses
on the investigation of gels containing nano-engines as in the experiments
of ref. \cite{Li2015} or the analysis of individual polymers attached
nano-engines in more complex environments than an a-thermal solvent.

\section{Summary}

In the present work, we have presented simulation data and a scaling
analysis of a figure of 8 shaped molecule (``T8'') where two polymer
loops are coiled up against each other by a Feringa engine that is
located at the core of the molecule. Our observations support Grosberg's
model \cite{Grosberg1996,Grosberg2000} for the physics of knotted
polymers. Conformation data can be overlapped when plotting it as
a function of $W_{\text{n}}N^{-1/4}$, where $W_{\text{n}}$ is the
winding number of the two strands that is proportional to the number
of blobs inside the ``knotted'' region of the molecule. The change
in the size of the molecules as well as the folding of the two strands
follows roughly scaling predictions that are derived by assuming that
the strands are confined inside an effective tube, in agreement with
the model assumptions in refs. \cite{Grosberg1996,Grosberg2000}.
Finally, our data supports a weak localization of the knots with localization
exponent $t\approx0.78$ in case of single polymers in the a-thermal
limit.

\section{Acknowledgement}

The authors acknowledge funding from the DFG grants LA 2735/5-1 and
SO 277/17 and a generous grant of computation time at the ZIH Dresden.

\section{Appendix}

\subsection{Computation of the change in the angle of twist \label{subsec:Computation-of-the}}

\begin{figure}
\begin{center}\includegraphics[width=0.6\columnwidth]{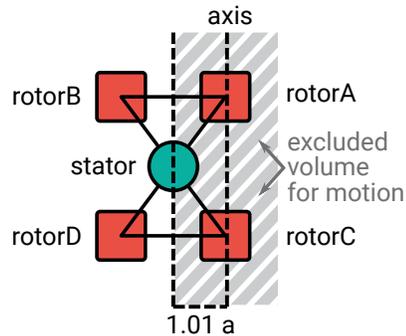}\end{center}

\caption{\label{AufbauWinkelz=0000E4hlung}Sketch of the engine, rotation axis,
and the volume that is excluded for the motion of rotor monomers B
and D.}
\end{figure}

The change in the angle of twist $\Delta\alpha$ of each motor is
computed each time one rotor monomer is moved in the simulation. Computation
requires particular care for lattice based simulations, where singularities
in the computation of the twist angle are rather frequent, for instance,
when a rotor is getting parallel to the rotation axis. Also, the large
jump size allows for a significant number of jumps that lead to angular
changes by $\pi$ (jumps across the axis), for which it is not clear
whether these refer to a left handed or right handed torsion. Furthermore,
the rotation axis needs to move with the engine, i.e. it must be defined
by monomers of the engine. A reorientation of the axis causes additional
changes in the angle of twist of the rotor that is stationary. The
net change in torsion angle with respect to moved and stationary rotor
needs to be restricted to less than $\pi$ in order to correctly assign
the handedness of the twist. All of these conditions are satisfied
by the following definition of the rotation axis in combination with
some additional restrictions on the motion of the rotor monomers.

The rotation axis $\overrightarrow{A}$ is defined as the positions
of rotor monomers A and C and pointing from C to A, 
\begin{equation}
\overrightarrow{A}=\overrightarrow{R}_{\text{A}}-\overrightarrow{R}_{\text{C}},\label{eq:A}
\end{equation}
as shown in figure \ref{AufbauWinkelz=0000E4hlung}. In order to assure
that the computation of the angles leads to no errors, the rotor monomers
B and D are forced to stay a minimum distance of more than one lattice
unit away from the rotation axis. Also, an attempted motion of an
axis monomer is rejected, if it leads to new positions that are in
conflict with this condition. This prevents that a rotor becomes parallel
to the axis and that jumps across the axis occur, since jumps are
of one lattice unit length. Because of this jump length, also the
maximum change in the angle of twist per rotor is restricted to less
than $\pi/3$ and the total change in twist remains below $2\pi/3$,
which allows to correctly assign the handedness.

For computation of the angle of twist we distinguish between the two
cases where the axis moves or where it is stationary. In the former
case, we can directly proceed to the computation of the angle of twist,
equations (\ref{eq:Z})-(\ref{eq:Delta_alpha}) while otherwise, we
have to first rotate the current coordinates into the future coordinate
system. To simplify notation, we add an index $a$ to specify the
new coordinates \emph{after} the move and an index $r$, if these
coordinates need to be considered in a \emph{rotated} coordinate system.
If the axis changes during the move, i.e. when monomer A or C move,
we first check whether the current axis $\overrightarrow{A}$ and
future axis $\overrightarrow{A_{\text{a}}}$ after the move are parallel.
If yes, the change in the angle of twist is zero and we are done.
If not, we compute the rotation matrix that rotates axis $\overrightarrow{A}$
into the new direction $\overrightarrow{A_{\text{a}}}$ such that
yaw, pitch and roll angles are correctly separated, and the change
in twist angle is solely related to the change in roll angle. To this
end, we first construct an orthonormal base from the the plane of
axis reorientation using 
\begin{equation}
\overrightarrow{N}=\overrightarrow{A}\times\overrightarrow{A_{\text{a}}}\label{eq:N}
\end{equation}
and
\begin{equation}
\overrightarrow{C}=\overrightarrow{N}\times\overrightarrow{A}.\label{eq:C}
\end{equation}
Next, we normalize all vectors $\overrightarrow{e_{\text{X}}}=\overrightarrow{X}/|\overrightarrow{X}|$
with $X=A,C,N$ and define the orientation matrix prior to the move,
\begin{equation}
\boldsymbol{M}_{p}=\left(\overrightarrow{e_{\text{C}}},\overrightarrow{e_{\text{N}}},\overrightarrow{e_{\text{A}}}\right).\label{eq:Mp}
\end{equation}
The orientation matrix after the move, $M_{\text{a}}$, is obtained
by using
\begin{equation}
\overrightarrow{N_{\text{a}}}=\overrightarrow{A_{\text{a}}}\times-\overrightarrow{A}\label{eq:N2}
\end{equation}
and
\begin{equation}
\overrightarrow{C_{\text{a}}}=\overrightarrow{N}\times\overrightarrow{A_{\text{a}}}\label{eq:C2}
\end{equation}
in similar manner as above from the normalized vectors $\overrightarrow{e_{\text{X}}}$:
\begin{equation}
\boldsymbol{M}_{\text{a}}=\left(\overrightarrow{e_{\text{\ensuremath{C_{a}}}}},\overrightarrow{e_{\text{\ensuremath{N_{a}}}}},\overrightarrow{e_{\text{\ensuremath{A_{a}}}}}\right).\label{eq:Ma}
\end{equation}
The rotation matrix $\boldsymbol{M}$ is then given by right multiplication
of the matrix after the move with the inverted matrix prior to the
move 
\begin{equation}
\boldsymbol{M}=\boldsymbol{M}_{\text{a}}\boldsymbol{M}_{\text{p}}^{-1}.\label{eq:M}
\end{equation}
This rotation matrix is used to rotate the upper

\begin{equation}
\overrightarrow{U_{\text{r}}}=\boldsymbol{M}\overrightarrow{U}=\boldsymbol{M}\left(\overrightarrow{R}_{\text{B}}-\overrightarrow{R}_{\text{A}}\right)\label{eq:U}
\end{equation}
and lower rotor 
\begin{equation}
\overrightarrow{L_{r}}=\boldsymbol{M}\overrightarrow{L}=\boldsymbol{M}\left(\overrightarrow{R}_{\text{D}}-\overrightarrow{R}_{\text{C}}\right).\label{eq:L}
\end{equation}
The rotated vectors $\overrightarrow{U_{\text{r}}}$ $\overrightarrow{L_{\text{r}}}$,
and $\overrightarrow{e_{\text{\ensuremath{A_{a}}}}}$ replace $\overrightarrow{U}$,
$\overrightarrow{L}$, and $\overrightarrow{e_{\text{A}}}$ below,
if the axis is moving. 

Due to the symmetry of the problem, we denote here only the change
in the angle of twist for the upper rotor. Obviously, $\overrightarrow{A}=\overrightarrow{A_{\text{a}}}$,
if the axis is not moving. The scalar product of $\overrightarrow{e_{\text{\ensuremath{A_{a}}}}}$
times $\overrightarrow{U}$ (or$\overrightarrow{U_{r}}$, if the axis
is moving) provides the vector component parallel to the axis $\overrightarrow{A_{\text{a}}}$,
which we use to compute the vector component 
\begin{equation}
\overrightarrow{O_{\text{U}}}=\overrightarrow{U}-\overrightarrow{A_{\text{a}}}\cdot\left(\overrightarrow{U}\cdot\overrightarrow{e_{\text{\ensuremath{A_{a}}}}}\right)\label{eq:Z}
\end{equation}
of the upper rotor orthogonal to axis $\overrightarrow{A_{\text{a}}}$.
The same computations are repeated for $\overrightarrow{U_{\text{a}}}$
and provide $\overrightarrow{O_{\text{\ensuremath{U_{a}}}}}$. The
angle between both orthogonal components of the upper rotor, $\alpha_{\text{U}}$
, is given by 
\begin{equation}
\alpha_{\text{U}}=\arccos\left(\frac{\overrightarrow{O_{\text{U}}}\overrightarrow{O_{\text{\ensuremath{U_{a}}}}}}{\left|\overrightarrow{O_{\text{U}}}\right|\left|\overrightarrow{O_{\text{\ensuremath{U_{a}}}}}\right|}\right).\label{eq:alpha}
\end{equation}
The chirality of the upper rotor, $c_{\text{U}},$ is one, $c_{\text{U}}=1$,
if 
\begin{equation}
\overrightarrow{A'}\cdot\left(\overrightarrow{O_{\text{U}}}\times\overrightarrow{O_{\text{U'}}}\right)>0,\label{eq:Chirality}
\end{equation}
which determines, whether the change in angle of twist of the upper
rotor is positive ($c_{\text{U}}=-1$ otherwise). If only the upper
rotor is moving, the change in the angle of twist is given by
\begin{equation}
\Delta\alpha=\alpha_{\text{U}}c_{\text{U}}.\label{eq:Delta_alpha-1}
\end{equation}
Note that the chirality of the lower rotor, $c_{\text{L}}$ is computed
analogously but with opposite sign in order to provide the twist of
upper rotor with respect to the lower rotor.

If the axis is moving, we compute the change in angle of twist for
both upper and lower rotor separately. The total change in angle of
twist is then
\begin{equation}
\Delta\alpha=\alpha_{\text{U}}c_{\text{U}}+\alpha_{\text{L}}c_{\text{L}}.\label{eq:Delta_alpha}
\end{equation}
This $\Delta\alpha$ is used to compute the potential energy difference
in equation (1) of the main manuscript.

\subsection{Contact analysis\label{subsec:Contact-analysis}}

In order to learn about the conformations of the T8, contacts between
segments of the two polymer loops were analyzed. The monomers of each
loop were labeled from $i=2$ to $N+1$ in order to compute the elastic
strand 
\begin{equation}
N_{\text{el}}(i)=\frac{i(N+2-i)}{N+2}\label{eq:Nel}
\end{equation}
to the center monomer of the engine similar to previous work on cyclic
polymers \cite{Lang2013}. $N_{\text{el}}$ is taken as a rough estimate
for the number of segments for an equivalent self-avoiding walk that
describes the return probability to the center of the T8, which refers
to the closest segments of the second loop. Due to the quick decay
of the return probability of inside chain contacts \cite{Redner1980,DesCloizeaux1980},
$P_{\text{c}}\propto N_{\text{el}}^{-(d+\theta_{2})\nu}\propto N_{\text{el}}^{-2.18}$,
we expect that this estimate captures the approximate scaling of the
contacts without applied torque. Note that we consider the contact
exponent $\theta_{2}$ for contacts between inner segments of chains,
as the T8 has no ends. Contacts are determined from simulation data
by analyzing a sphere with radius 3 lattice units around a given monomer.
The contact probability $P_{\text{c}}$ is defined by the event that
at least one monomer of the other loop resides within this sphere
and it is averaged over a long time series of conformations.

\begin{figure}
\includegraphics[angle=270,width=1\columnwidth]{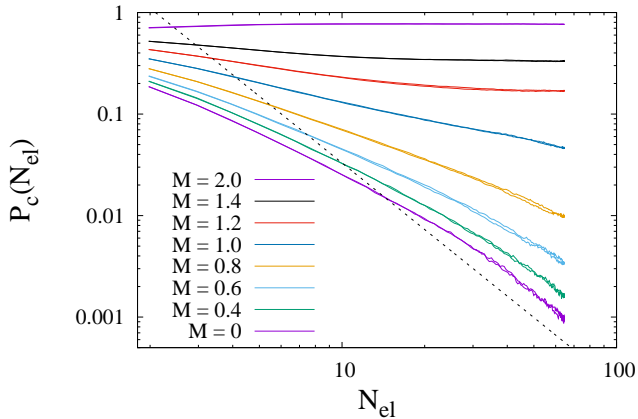}

\caption{\label{fig:Contacts-between-segments}Contacts between segments of
the loop as a function of the elastic strand $N_{\text{el}}$ to the
center monomer of the engine for a T8 with $N=256$. The thin dotted
line indicates a scaling $P_{\text{c}}\propto N_{\text{el}}^{-(d+\theta_{2})\nu}\propto N_{\text{el}}^{-2.18}$.}
\end{figure}
The results of this analysis are plotted in Figure \ref{fig:Contacts-between-segments}
for some selected torques. The obvious trend is that contacts increase
with increasing torque. In the limit of low torques, $M\approx0$,
the contact probability follows approximately the scaling expected
for a self-avoiding walk for large $N_{\text{el}}$. This indicates
that the simplifications above and the mapping on $N_{\text{el}}$
are consistent with the simulation data. Saturation $P_{\text{c}}\approx const.$
is only reached for rather high torques $M\gtrsim1.2$ where at least
1 out of 5 monomers is in contact with a monomer of the second loop.
Note that the winding numbers for $M=0.6,\ 0.8,\text{\ 1.0,}$ and
$1.2$ are $W_{\text{n}}\approx0.99$, $1.99$, $4.68$, and $10.7$.
Since there are about 3.7 blobs per winding number, the coils start
to be overstretched already at torques $M\gtrsim1$ close to the qualitative
change of the contact statistics. The data between these cases shows
a gradual transition between both asymptotic limits, whereby at larger
$N_{\text{el}}$ an increased downturn of $P_{\text{c}}$ is visible,
which indicates the preferred localization of entanglements (and thus,
enhanced contacts) next to the engine. 

\begin{figure}
\includegraphics[angle=270,width=1\columnwidth]{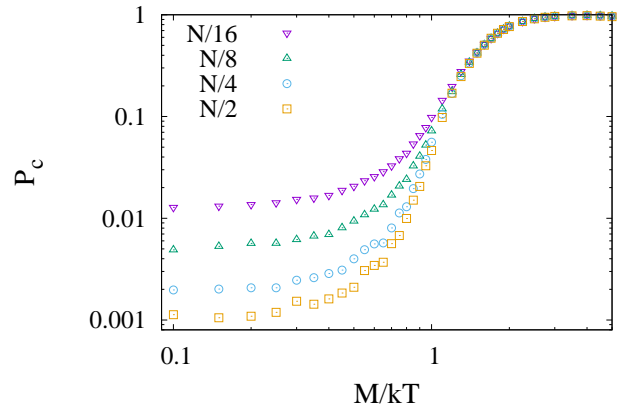}

\caption{\label{fig:Contact-probability-}Contact probability $P_{c}$ between
loops for selected monomers at $i=N/2$, $N/4$, $N/8$, and $N/16$.
Data was recorded for the T8 with $N=256$.}
\end{figure}

An alternate view on the contact statistics is provided in Figure
\ref{fig:Contact-probability-}, where $P_{\text{c}}$ is analyzed
as a function of the applied torque for a given segment index $i$.
Folded conformations show up here by a collapse of the data at same
torque as we observe for $M/kT\gtrsim1.2$. The missing collapse of
the data at low torque points towards a localization of the entanglements
next to the engine of the tanglotron.

\subsection{Numerical correction to localization\label{subsec:Numerical-correction-to}}

The determination of the size of the knotted region is a non-trivial
task, which is in most cases performed by an algorithm of the type
described in \cite{Grosberg2016}: ``... usually by the steepest
descent, the smallest among the spheres'' is selected, ``which satisfy
two properties: first, the sphere has to be pierced by the polymer
exactly two times (thus allowing for an unambiguous determination
of the topology for the inside section of the polymer); second, the
portion of the polymer inside of the sphere is topologically equivalent
to the entire chain (which means the sphere encloses the knot). Polymer
length inside this sphere is $L_{\text{knot}}$.'' 

In order to understand whether such an algorithm leads to some numerical
bias for the weight fraction 
\begin{equation}
w_{\text{knot}}=N_{\text{knot}}/N=L_{\text{knot}}/(bN)\label{eq:wknot}
\end{equation}
of polymer inside the knot or not, let us perform the following gedankenexperiment:
we consider a cyclic polymer where we randomly place $l$ labels among
the $N$ segments of the chain. These labels regard randomly selected
sections at which the polymer may wind around a second section of
the chain and thus, could be the position where the polymer may cut
through a minimum sphere as for the above algorithm. The largest among
the sections between two subsequent labels will be chosen as the unknotted
part of the polymer. Note that the case of a random choice of sections
refers to zero localization of the ``knot''. The question that we
would like to answer is whether placing the labels along the chains
leads to a statistical bias and thus, also for the algorithm for determining
the weight fraction of the knot.

\begin{figure}
\includegraphics[angle=270,width=1\columnwidth]{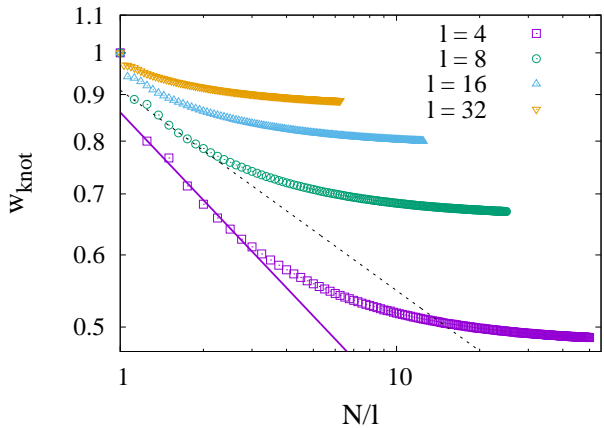}

\caption{\label{fig:Estimated-weight-fraction}Estimated weight fraction of
a knot $w_{\text{knot}}$ for a set of random labels along a chain.
The continuous line is a power law fit to the small $N/l$ data for
$l=4$ with exponent $-0.32\pm0.01$ and the dashed line indicates
a power of $-0.22$.}
\end{figure}

The result of this gedankenexperiment are shown in Figure \ref{fig:Estimated-weight-fraction}.
We observe a monotonously decaying $w_{\text{knot}}$ for increasing
$N$ at constant $l$ that might be interpreted as a weak localization
of a knot despite of the fact that the labels were placed randomly.
For sufficiently large $N/l$, this correction can be ignored. Note
that an analytical description of this gedankenexperiment can be obtained
by adapting the computations in ref. \cite{Lang2003b} (first label
can be considered as to cut the cyclic polymer in a linear one).

Let us use the data in Figure \ref{fig:Estimated-weight-fraction}
for a rough estimate of the minimum required $N$ to observe unbiased
results for the example of a trefoil knot. Apparently, the numerical
correction starts to saturate roughly for $N/l\apprge10$. When adopting
a scaling picture as proposed by Grosberg \cite{Grosberg1996,Grosberg2000},
we identify the number $N$ with the number of minimal units, which
is confinement or tension blobs of the chain. $l$ cannot be smaller
than the minimum number of intersections when projecting the knot
to a plane. Thus, the chains should consist of at least 30 blobs for
an unbiased result. Furthermore, the polymer strands inside the blobs
should not be overstretched to observe the proper scaling of the data.
This requires, that blob size $\xi$ as compared to fully stretched
size of the corresponding chain section, $bg$, fulfills the condition
$\xi/(bg)\lesssim1/5$ \cite{Rubinstein2005}. Therefore, $750$ segments
for a self-avoiding cyclic polymer are just a lower bound for the
required minimum number of chain sections to observe a correct scaling
of knot localization. This criterion should be surpassed by at least
one order of magnitude to determine the localization exponent, which
has not been met in previous work. Recall also that the measured knot
size may show additional dependencies on the algorithm used \cite{Tubiana2011b}.
These problems are avoided by our approach as we measure the confining
force within the very first blob. 

\bibliographystyle{achemso}
\bibliography{Quellen}

\end{document}